*Article*

# Quantum-Optical Spectrometry in Relativistic Laser–Plasma Interactions Using the High-Harmonic Generation Process: A Proposal

Theocharis Lamprou [1,2], Rodrigo Lopez-Martens [3,4], Stefan Haessler [3], Ioannis Liontos [1], Subhendu Kahaly [4], Javier Rivera-Dean [5], Philipp Stammer [5,6], Emilio Pisanty [6], Marcelo F. Ciappina [5,7,8,9], Maciej Lewenstein [5,10] and Paraskevas Tzallas [1,4,*]

1. Foundation for Research and Technology-Hellas, Institute of Electronic Structure & Laser, GR-70013 Heraklion, Greece; tlamprou@physics.uoc.gr (T.L.); iliontos@iesl.forth.gr (I.L.)
2. Department of Physics, University of Crete, GR-71003 Heraklion, Greece
3. Laboratoire d'Optique Appliquée, Institut Polytechnique de Paris, ENSTA-Paris, Ecole Polytechnique, CNRS, CEDEX, 91120 Palaiseau, France; rodrigo.lopez-martens@ensta-paris.fr (R.L.-M.); stefan.haessler@ensta-paris.fr (S.H.)
4. ELI-ALPS, ELI-Hu Non-Profit Ltd., Wolfgang Sandner utca 3., H-6728 Szeged, Hungary; subhendu.kahaly@eli-alps.hu
5. ICFO—Institut de Ciencies Fotoniques, The Barcelona Institute of Science and Technology, 08860 Castelldefels (Barcelona), Spain; Javier.Rivera@icfo.eu (J.R.-D.); philipp.stammer@gmail.com (P.S.); Marcelo.Ciappina@alumni.icfo.eu (M.F.C.); maciej.lewenstein@icfo.eu (M.L.)
6. Max Born Institute for Nonlinear Optics and Short Pulse Spectroscopy, Max Born Strasse 2a, D-12489 Berlin, Germany; emilio.pisanty@alumni.icfo.eu
7. Institute of Physics of the ASCR, ELI-Beamlines Project, Na Slovance 2, 182 21 Prague, Czech Republic
8. Physics Program, Guangdong Technion—Israel Institute of Technology, Shantou 515063, China
9. Technion—Israel Institute of Technology, Haifa 32000, Israel
10. ICREA, Pg. Lluís Companys 23, 08010 Barcelona, Spain
*  Correspondence: ptzallas@iesl.forth.gr

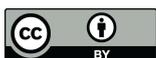

**Abstract:** Quantum-optical spectrometry is a recently developed shot-to-shot photon correlation-based method, namely using a quantum spectrometer (QS), that has been used to reveal the quantum optical nature of intense laser–matter interactions and connect the research domains of quantum optics (QO) and strong laser–field physics (SLFP). The method provides the probability of absorbing photons from a driving laser field towards the generation of a strong laser–field interaction product, such as high-order harmonics. In this case, the harmonic spectrum is reflected in the photon number distribution of the infrared (IR) driving field after its interaction with the high harmonic generation medium. The method was implemented in non-relativistic interactions using high harmonics produced by the interaction of strong laser pulses with atoms and semiconductors. Very recently, it was used for the generation of non-classical light states in intense laser–atom interaction, building the basis for studies of quantum electrodynamics in strong laser–field physics and the development of a new class of non-classical light sources for applications in quantum technology. Here, after a brief introduction of the QS method, we will discuss how the QS can be applied in relativistic laser–plasma interactions and become the driving factor for initiating investigations on relativistic quantum electrodynamics.

## 1. Introduction

A few years after the pioneering invention of lasers by Maiman [1], the scientific community addressed the two following fundamental questions: (I) "What is the quantum description of a classically oscillating current?" and (II) "How can we increase the laser power in order to observe the nonlinear response of matter?". Investigations dedicated to answering these questions led to the development of the research areas of quantum optics (QO) [2,3] and strong laser-field physics (SLFP) concerning interactions of matter with intense electromagnetic fields [4–11]. In QO, the quantum description of a classically oscillating electromagnetic field [2,3] opened the way for fascinating investigations in quantum technology [11–14]. Central to these studies are non-classical light sources [11,15–19] as they offer a unique resource for quantum communication, information and computation, as well as atomic physics, visual science, and high precision interferometry applied for the detection of gravitational waves, to name a few [11–14]. On the other hand, in SLFP, the development of high-power femtosecond (*fs*) pulsed laser sources [4,5,11] and of the classical and semi-classical description of intense laser-matter interaction [6–11,20–22] opened the way for studies ranging from relativistic electron acceleration [23,24], and high-order harmonic generation (HHG), to attosecond science and ultrafast optoelectronics [6–11,25].

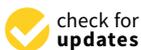

Despite the progress achieved in these research domains, they have remained disconnected over the years leaving the advantages emerging from the synthesis of both the QO and SLPF areas thus far unexplored. This is because, on the one hand, most of the studies in QO are performed using weak electromagnetic fields (low photon number light sources) where the interaction is described by fully quantized approaches treating the field quantum mechanically and affected by the interaction. Meanwhile, in SLFP, due to the high photon number of the driving laser field, the overwhelming majority of investigations rely on semi-classical approaches which treat the electromagnetic field classically and, therefore, as unaffected by the interaction. Recently, we have shown that QO and SLFP can be connected and used for investigations of strong laser-field quantum electrodynamics and the development of a new class of non-classical light sources [26]. Central to these studies is the implementation of a photon-correlation-based quantum spectrometer (QS) method. This method has been applied in non-relativistic interactions utilizing the high-order harmonics generated by the interaction of atoms [27] and semiconductors [28] with intense ($I_L < 10^{15}$ W/cm$^2$) infrared (IR) femtosecond (*fs*) laser fields. Taking into account the back action of the HHG process on the coherent quantum state of the driving field, the method provides the spectrum of the high harmonics by simply measuring the photon number distribution of the IR driving field after HHG, which directly reflects the probability of absorbing IR photons towards HHG. Here, after a brief introduction of the QS method (Section 1), we will discuss how the approach can be extended to interactions in the relativistic intensity region ($I_L > 10^{18}$ W/cm$^2$) using a currently available state-of-the-art laser-plasma HHG source [29] (Section 2). The universal application of QS may serve as the building block for the foundation of a new research direction that benefits from the synergy of QO and SLFP.

## 2. Operation Principle of the Quantum Spectrometer

The quantum spectrometer is a shot-to-shot photon correlation-based method which provides the probability of absorbing photons from a driving laser field towards the generation of a strong laser-field interaction product. Here, we briefly describe the fundamentals of the QS, considering the high harmonics generated in atoms as an interaction product. The operation principle of the QS is illustrated schematically in Figure 1. A p-polarized *fs* IR pulse with mean photon number $N_0$ (typically > $10^{14}$ photons per pulse) is focused into the laser–matter interaction region where the high harmonics are generated. The shot-to-shot energy fluctuations of the IR field are measured by means of an IR photodiode $PD_0$. The intensity of the IR pulse in the interaction region is typically in the range of $I_L \sim 10^{14}$ W/cm$^2$. After the interaction, the photon number of the generated extreme ultraviolet (XUV) harmonics and the IR field are $N_{XUV}$ and $N_{IR}$, respectively, with $N_{IR} < N_0$ due to IR photon losses associated with all processes taking place in the interaction region. The generated harmonics are separated from the IR beam by means of a harmonic separator (HS) plate placed at Brewster's angle for the p-polarized IR beam. The HS plate transmits the IR field and reflects the harmonics towards an XUV photon detector

PD$_{XUV}$. The photon number of the IR beam passing through the HS is measured by the IR photodiode PD$_{IR}$.

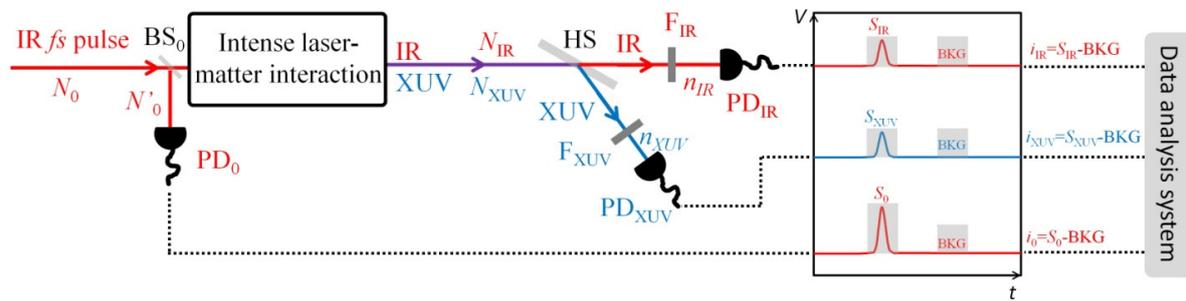

**Figure 1.** A schematic of the operation principle of a quantum spectrometer (QS). An intense *fs* infrared (IR) pulse with mean photon number $N_0$ is focused into the intense laser-matter interaction region where high harmonics are generated in the extreme-ultraviolet (XUV) spectral region. $N_{IR}$ (with $N_{IR} < N_0$) and $N_{XUV}$ are the photon numbers of the IR and XUV fields exiting the high-order harmonic generation (HHG) medium. HS: harmonic separation optical element used to separate the XUV and IR beam. The generated harmonics can be spectrally resolved by steering the harmonic beam towards a conventional XUV spectrometer (not shown here). PD$_0$: an IR photodiode used for measuring the shot-to-shot energy fluctuations of the IR driving pulse. BS$_0$: IR beam separator, used to reflect a small portion of the IR beam, with photon number $N'_0$, to the PD$_0$. PD$_{IR}$: An IR photodiode. PD$_{XUV}$: an XUV photodetector. $n_{IR}$ and $n_{XUV}$ are the IR and XUV photon number reaching the PD$_{IR}$ and PD$_{XUV}$, respectively. F$_{IR}$: IR neutral density filter used to attenuate the IR beam and avoid saturation effects on PD$_{IR}$. F$_{XUV}$: XUV thin metal filter used to select part of the HHG spectrum and eliminate any residual part of the IR beam. The V(*t*) graph illustrates the voltage signals of the PD$_{IR}$, PD$_{XUV}$ and PD$_0$. The gray shaded areas in this graph depicts the gates used by a boxcar integrator to record for each shot the photocurrents $S_{IR}$, $S_{XUV}$, $S_0$ of the PD$_{IR}$, PD$_{XUV}$, PD$_0$ and the corresponding background electronic noise (BKG). After the BKG subtraction photocurrents $i_{IR}$, $i_{XUV}$, $i_0$ are recorded and analyzed.

It is noted that the photodiodes PD$_0$ and PD$_{IR}$ should be identical. In order to avoid saturation effects, the IR field before reaching PD$_{IR}$ should be attenuated by a factor $T_{IR}$ by means of a neutral density IR filter F$_{IR}$. Also, a thin metal filter F$_{XUV}$ (typically aluminum of thickness in the range of few-hundred nm) transmits high harmonics and eliminates any residual part of the IR beam that is reflected by the HS. The $n_{IR}$ and $n_{XUV}$ photons reaching the PD$_{IR}$ and PD$_{XUV}$, respectively, are related with $N_{IR}$ and $N_{XUV}$ via the equations $n_{IR} = N_{IR}/(T_{IR}^{(HS)} T_{IR})$ and $n_{XUV} = N_{XUV}/(R_{XUV}^{(HS)} T_{XUV})$, where $T_{IR}^{(HS)}$, $T_{XUV}$ and $R_{XUV}^{(HS)}$ are the IR attenuation introduced by the HS, the XUV attenuation introduced by the metal filter, and XUV reflectivity of the HS plate, respectively. The $n_{IR}$, $n_{XUV}$ and $N_0$ ($N_0 = N'_0/R_{IR}$ where $R_{IR}$ is the reflectivity of the BS$_0$) are recorded for each laser shot by a high dynamic range boxcar integrator together with the corresponding background (BKG) electronic noise which is subtracted from the corresponding photon signal $S_{IR}$, $S_{XUV}$ and $S_0$ (panel in Figure 1). The photo current outputs $i_{IR} = S_{IR} - BKG$, $i_{XUV} = S_{XUV} - BKG$ and $i_0 = S_0 - BKG$ are analyzed by a data acquisition system. The data analysis procedure is divided in three steps: (I) To avoid any potentially strong fluctuations of the energy of the driving pulse, we use the $i_0$ and we collect the laser shots which provide high energy stability which is typically in the level of $\lesssim$1% i.e., we collect only the laser shots which provide a value close to the peak of the shot-to-shot energy distribution, (II) after balancing the mean value of $i_{XUV}$ to the mean value of $i_{IR}$, we create the joint XUV-vs.-IR photon number distribution as is shown with gray points in the ($i_{IR}$, $i_{XUV}$) plot shown in Figure 2a. This distribution contains information about all the processes taking place in the interaction region-including those that are irrelevant to the harmonic process, and (III) based on energy conservation (when XUV photons are generated the IR field loses photons), we select only the shots that provide a signal along the anti-correlation diagonal (red points in Figure 2a).

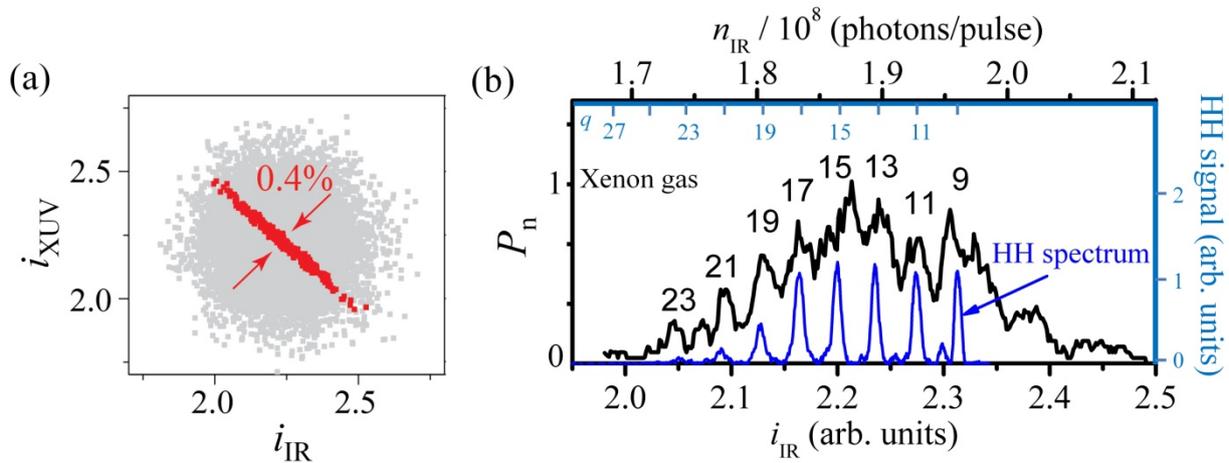

**Figure 2.** (**a**) Gray points: Joint IR-vs.-XUV photon number distribution. The high harmonics were generated by the interaction of an ≈ 25 fs IR laser pulse with Xenon gas. Red points: points along the anti-correlation diagonal of width $w_{anti} \approx 0.4\%$. (**b**) Black line: probability of absorbing IR photons ($P_n$) towards the harmonic emission. The continuum is a background of the XUV-IR uncorrelated shots introduced by the QS, due to a larger width than the optimum value of the anti-correlation diagonal, as explained in more detail in the text. Blue line: High harmonic spectrum recorded by a conventional grating XUV spectrometer. The vertical and horizontal axis of the HH spectrum are shown in blue color. The figures are reproduced from ref. [27].

This is the optimal and physically acceptable way to collect only the shots that are relevant to the HHG and remove the unwanted background associated with all processes irrelevant to the harmonic generation. The width of the anti-correlation diagonal, which is associated with the resolution of the QS, is determined by the accuracy to find the peak of the joint distribution i.e., $w_{anti} = W_{joint}/k^{1/2}$, (where $W$ is the percentage of the width of the $i_{IR}$ relative to its mean value, and $k$ is the number of points in the XUV-vs-IR distribution). The points along the anti-correlation diagonal provide the probability of absorbing IR photons ($P_n$) towards the harmonic emission [28,30]. This is shown in Figure 2b. The $P_n$ (black line in Figure 2b) depicts a multi-peak structure corresponding to the generated high harmonic orders. For comparison reasons, the spectrum of the high harmonics as recorded by a conventional XUV grating spectrometer is shown in blue in Figure 2b.

The implementation of the QS method and its extension to relativistic laser–plasma interactions depends on the power of the QS to resolve the multi-peak structure of $P_n$ in the case of relativistic HHG. This is associated with: (a) the dynamic range of the detection system and with the number of accumulated shots i.e., the higher the number of accumulated shots the better the resolution, and (b) the number of IR photons absorbed towards the harmonic emission (associated with the conversion efficiency) which is connected with the IR photon number difference between consecutive peaks in the multi-peak structure of $P_n$ ("peak spacing" on $P_n$) i.e., the higher the conversion efficiency of the HHG process the higher the "peak spacing" on $P_n$. Considering that the number of IR photons absorbed towards the emission of the $q$th harmonic is $N_q^{(IR)} = A \cdot q \cdot N_q^{(XUV)}$ (where $N_q^{(XUV)}$ is the photon number of the $q$th harmonic after exiting the harmonic generation medium and $A > 1$ is the absorption factor of the harmonic due to propagation in the medium), the "peak spacing" in $P_n$, for $\Delta q = 1$ (in case of generating odd and even harmonic orders), is expected to be $\Delta N_q^{(IR)} = A \cdot N_q^{(XUV)}$ corresponding to an $\Delta n_q^{(IR)} = \left(A \cdot N_q^{(XUV)}\right)/(T_{IR}^{(HS)} T_{IR})$ on the $PD_{IR}$ detector. Thus, in order to minimize the background associated with the IR-vs.-XUV uncorrelated shots and resolve the multi-peak structure in $P_n$, the width of the anti-correlation diagonal should be $w_{anti} = W_{joint}/k^{\frac{1}{2}} \lesssim \Delta n_q^{(IR)}/n_{IR}$, resulting in $w_{anti} \lesssim A \cdot N_q^{(XUV)}/N_{IR}$.

To provide an example, let us assume that we use a *fs* IR pulse of $N_0 \sim 10^{15}$ photons for the generation of high harmonics. Here, for reasons of simplicity we make the assump-tion that $N_{IR} \sim N_0$, the power of non-linearity of the HHG process is ~4–7, and the joint distribution contains $k \sim 10^6$ shots. Under these conditions, $W_{joint} \sim 6\%$ and $w_{anti} \sim 6 \times 10^{-3}\%$. Using the relation $N_q^{(XUV)} \gtrsim N_{IR} \cdot w_{anti}/A$, the latter denotes that the QS can resolve a background free multi-peak structure of $P_n$ when the number of XUV photons outgoing from the medium is $N_q^{(XUV)} \gtrsim 10^{11}/A$ photons per pulse and the dynamic range of the detection system is $D > \frac{N_{IR}}{A \cdot N_q^{(XUV)}} = 10^4/A$ i.e., $D > 14$ bit for $A = 1$. This level of harmonic photon numbers is typically generated in the majority of HHG experiments [31]. It is noted that for interactions where the XUV radiation propagates in the HHG medium (such as gases and crystals), $A >> 1$ (typically in the order of ~$10^3$ depending on the medium and wavelength of the XUV radiation) which significantly reduces the required XUV photon number $N_q^{(XUV)}$. To this end, we note that for practical reasons, in order to increase the visibility and the statistical significance of the multi-peak structure of $P_n$ while keeping the background at a negligible level, we increase the number of shots along the anticorrelation diagonal by increasing the width of this anticorrelation diagonal i.e., practically the optimal value of the width ($w_{anti}^{(opt.)}$) which provides the best visibility is in the range $w_{anti} < w_{anti}^{(opt.)} < \Delta n_q^{(IR)}/n_{IR}$. In the case where $w_{anti} \sim \Delta n_q^{(IR)}/n_{IR}$, the multi-peak structure of $P_n$ can also be obtained, but in this case the structure will build on the top of a non-negligible background IR photon number distribution which, depending on the application, has to be evaluated. For example, in case of using carrier-envelope-phase (CEP)-controlled few-cycle driving pulses, quantitative information about the background contribution can be obtained by firstly using a sinusoidal laser field (or a multi-cycle laser field) and retrieving the multi-peak structure $P_n$, and then moving to the interaction with a cosine few-cycle laser field where an XUV continuum spectrum is generated.

In the following section we describe how the QS can be implemented in relativistic laser–plasma interactions using the advantages provided by the unique properties of the currently operational high repetition rate laser-plasma high harmonic source [29].

## 3. Quantum Spectrometry in Laser–Plasma Interactions

The process of relativistic high harmonic generation (RHHG) [11,31–39], induced by the interaction of intense *fs* laser pulses with solid surfaces (with $I_L \cdot \lambda^2 > 1.37 \times 10^{18}$ W/cm$^2 \cdot \mu$m$^2$, such that the normalized vector potential $a_0 = \left( I_L[\text{W} \cdot \text{cm}^{-2}] \times \frac{\lambda_L^2[\mu\text{m}^2]}{1.37 \times 10^{18}} \right)^{1/2} > 1$, where $\lambda$ is the carrier wavelength of the driving laser pulse), is one of the most suitable and commonly used ways for accessing interactions in the relativistic regime. Under appropriate conditions, the process can be described by the relativistic oscillating mirror (ROM) model [32–34] or the relativistic electron spring/coherent synchrotron emission models [35,36]. Early one-dimensional numerical simulations [40] predicted that using ~5 *fs* IR laser pulses with intensity on target in the range of $I_L \sim 10^{20}$ W/cm$^2$, high harmonics in the photon energy range of ~60 eV can be generated with conversion efficiency in the range of $10^{-1}$–$10^{-2}$, while the generation of harmonics in the keV photon energy range can take place with conversion efficiency in the range of ~$10^{-4}$. More recent two-dimensional numerical simulations [29,41–43] for $I_L \sim 10^{19}$ W/cm$^2$ predict typically two orders of magnitude lower efficiencies, in agreement with experiments [42–44], i.e., ~$10^{-4}$ in the range of 30–70 eV.

Employing relativistic high-harmonics (RHH) (with which we will designate harmonics generated in the relativistic regime in general, including those generated via the ROM mechanism) for implementing the QS method poses several challenges that can be circumvented with clever experimental design. Firstly, in order to achieve high intensity in the interaction region, a prerequisite for RHH, a tight focusing condition is necessary. Furthermore, the RHH generation efficiency is optimized by tuning the plasma density gradient [45,46]. These two criteria together, increase the resulting high harmonic beam divergence making the beam collection and manipulation an issue to be addressed. Nevertheless, it has been experimentally demonstrated that the effective RHH beam divergence, can be reduced by controlling the laser spatial phase [47]. Secondly, the relativistic high harmonic attosecond pulse train from surfaces is also accompanied by concomitant generation of co-propagating energetic electron bunches [48], an additional interaction end-product complicating the energy partitioning for the QS. However, recent experiments have demonstrated the simultaneous measurement of both the electron and high harmonic signals unraveling a quantitative correlation between the harmonic and relativistic electron emissions for short-gradient scale lengths [41] relevant to RHH. Thirdly, the foundation of QS is based on high statistics which implies that high repetition rate lasers should be used for RHHG. Therefore, the simultaneous requirement of high intensity for interaction and high repetition rate for implementation confines the choice of the laser to the few-cycle domain. This necessitates CEP-controlled operation, since in this domain CEP induced changes in the HHG spectra [29,44,49]. All these factors are taken into account in the current proposed scheme.

An ideal RHHG source, which can approach the aforementioned theoretical predictions and can, therefore, be used for implementing the QS method, has been recently demonstrated in reference [29]. The uniqueness of this source relevant for hosting the QS method, relies on: (a) the high repetition rate (1 kHz); (b) the CEP controlled few-cycle main pulse which generates the harmonics; (c) the high stability of the HHG process, associated with the stability and the flatness of the solid target; (d) the amount of the available shots which can be up to $10^6$ shots per target; (e) the control of the plasma gradient achieved by means of a pre-pulse with adjustable delay; and (f) the high conversion efficiency of the relativistic HHG process.

A schematic drawing of the source is shown in the right panel of Figure 3a. In short (details can be found in reference [29]), the HHG process is driven by a p-polarized $\approx$3.6 *fs* CEP controlled laser pulse of carrier wavelength $\approx$750 nm and photon number $N_0 \approx 10^{16}$ per pulse (i.e., of energy $\approx$ 2.6 mJ per pulse with $\approx$1% shot-to-shot energy fluctuation). The laser pulse is tightly focused (<2 μm full width at half maximum (FWHM) focal spot diameter shown in Figure 3a) by an *f*/1.5, 30° off-axis parabola onto the fused silica target at an incidence angle of $\theta \approx 55°$. The intensity of the focused pulse on the surface of the target is $I_L \approx 10^{19}$ W/cm$^2$ corresponding to $a_0 \approx 2$ enabling the generation of RHH.

The laser–plasma interaction is controlled by means of a spatially separated pre-pulse (delayed by few ps) which prepares the optimal plasma density gradient for RHHG. It is worth noting that the pre-pulse beam can be arranged to be non-collinear with the main pulse i.e., the pre-pulse to be reflected from the target at a different angle compared to the main pulse and the generated XUV beam. The polarization of the pre-pulse can be made orthogonal to that of the main pulse without affecting the RHHG process. The pre-pulse, created by picking off and then recombining $\approx$ 4% (corresponding to a photon number $N_{\text{pre-pulse}} \approx 4 \times 10^{14}$ photons per pulse) of the main pulse through holed mirrors, is focused to a much larger $\approx$ 13 μm FWHM spot (shown in gray scale in Figure 3a) in order to generate a transversally homogeneous plasma layer expanding into vacuum. The plasma density scale length *L* was controlled by setting a delay of $\approx$ 2 ps between the pre- and the main pulse. This results to an $L \approx \lambda/20$ (measured by mean spatial-domain interferometry [51]) which optimizes the conditions for RHH harmonic emission [45,46]. In Figure 3b we show a representative RHH spectrum generated by this source using a sinusoidal (CEP = $\pi/2$) $\approx$ 3.6 fs driving field. At these conditions, taking into account Figure 3c and the values obtained in the work of references [42–44], the conversion efficiency of the HHG process (Figure 3c) can roughly be considered to be in the range of $\eta_{\text{XUV}} \sim 10^{-4}$–$10^{-3}$, resulting in $N_q^{(XUV)} = (\eta_{XUV} N_0)/10 \sim 10^{11}$–$10^{12}$ photons per harmonic per pulse (considering, for reasons of simplicity, that the XUV radiation contains 10 harmonics of equal amplitudes). It is noted that this value is obtained for XUV photon energies in the range of 20–70 eV (i.e., central photon energy $\approx$ 45 eV). Taking into account a dependence of the harmonic yield on the harmonic order $\eta_q \propto \omega_q^{-8/3}$ it can be estimated that the conversion efficiency in the spectral range of the measured harmonic spectrum shown in Figure 2b (i.e., central photon energy $\approx$ 17 eV) can be $\eta_{\text{XUV}} \sim 10^{-3}$–$10^{-2}$, resulting to $N_q^{(XUV)} \sim 10^{12} - 10^{13}$ photons per harmonic per pulse. The stability of the CEP (measured by an *f*-2*f* spectral interferometer) is in the range of ≈500 mrad, a value which provides a sufficiently precise control of the sub-cycle HHG process. The highly stable rotating target holder [50] keeps the target precisely on the IR focus position (along the propagation axis) and thus secures the shot-to-shot intensity stability of the interaction. It is found that all these fluctuations, including the fluctuations of other parameters such as pulse duration and energy, are very small (<1%) and do not practically affect the HHG process. Finally, taking into account the aforementioned properties of the source and the ≈20 cm diameter disc-shape target, it can be estimated that the use of a single target can deliver up to ~$10^6$ highly stable shots of RHH (having the spectrum shown in Figure 3c) with $N_q^{(XUV)} \sim 10^{12} - 10^{13}$ photons per harmonic at 1 kHz repetition rate.

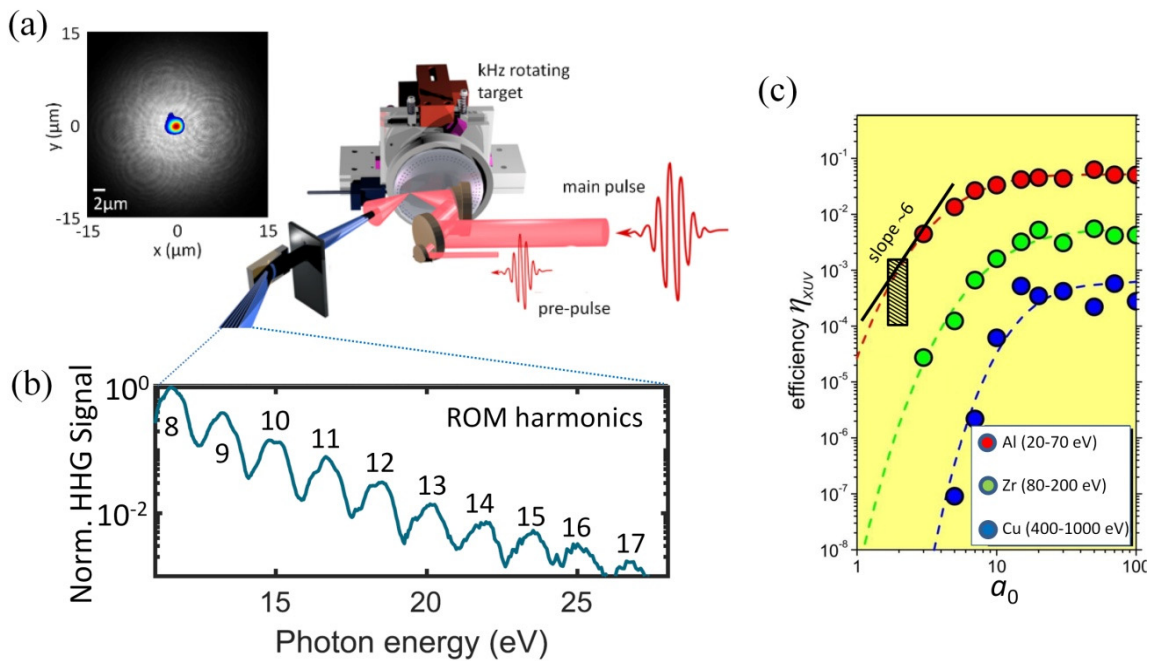

**Figure 3.** (**a**) A drawing of the 1 kHz repetition rate relativistic high harmonic generation (RHHG) source. A p-polarized ≈ 3.6 *fs* carrier-envelope-phase (CEP) stablilized IR pulse is focused onto a rotating $SiO_2$ disc target. The <2-μm, <100-μrad target surface stability during the rotation is ensured by precise alignment and high-quality bearings [50], ensuring the intensity stability in each laser shot. The number of available shots on the target is ~$10^6$. The plasma expansion is initiated by a pre-pulse arriving on the target a few ps before the main pulse. The spatial profiles of the pre- and main pulses on the target surface are shown on the left panel in gray and color scale, respectively. The spot size of the pre-pulse is much larger than the main pulse in order to ensure a homogenous plasma expansion. Spatial filtering approaches can also be used for optimizing the transmission/detection of the RHH and eliminate any residual part of the XUV radiation generated by any contribution of the coherent wake emission HHG process. (**b**) The spectrum of the generated RHH (generated by a sinusoidal 3.6 fs field) is recorded by an angle-resolved conventional XUV grating spectrometer. (**c**) Calculated conversion efficiency of the HHG process for different values of $a_0$ and different XUV spectral regions. The square line-shaded area depicts a rough value of the conversion efficiency estimated for the conditions used for the generation of the harmonics shown in (**b**). Using this figure we can roughly estimate, that the non-linearity of the RHH emission with respect to the driving field intensity in the region of $a_0 \approx 2$ is ~6 (solid black line). The figures (**a**) and (**b**) are reproduced from ref. [29], and the (**c**) from refs [31,40].

The proposed arrangement for implementing the QS method in the RHH beam-line is sketched in Figure 4a. It is a combination of the RHH beam line described above, and the QS described in Section 1. The definition of the symbols of HS, $F_{XUV}$, $F_{IR}$, $PD_0$, $PD_{IR}$, $PD_{XUV}$, $i_{IR}$, $i_{XUV}$, $i_0$, $n_{IR}$ and $n_{XUV}$, in Figure 4a are the same as those used in Section 1. In this configuration, the p-polarized ≈ 3.6 *fs* CEP-stabilized IR pulse will be focused onto a rotating disc-shape solid target. The plasma expansion will be initiated by an s-polarized pre-pulse arriving at the target a few-ps before the main pulse at a slightly different angle of incidence than the main pulse. The shot-to-shot energy fluctuation of the main pulse will be measured by the photodiode $PD_0$ (providing the photocurrent value $i_0$).

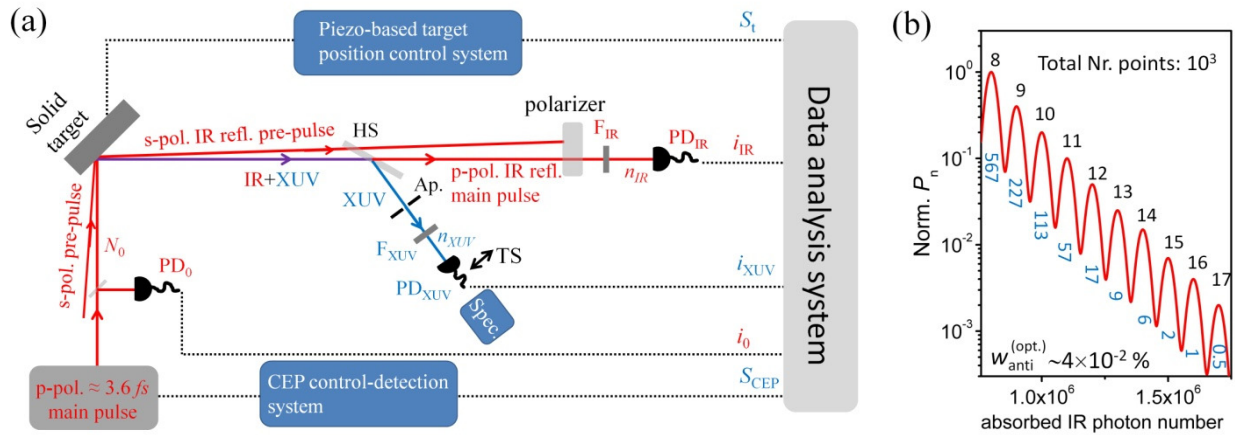

**Figure 4.** (**a**) A drawing of the proposed experimental set-up based on the operation principle of the QS described in Section 2, and the RHH beam line. A p-polarized ≈ 3.6 *fs* CEP stable IR pulse of $N_0 \sim 10^{16}$ photons per pulse is focused onto rotating disc shape solid target. The plasma expansion is initiated by an s-polarized pre-pulse arriving on the target few-ps before the main pulse and at slightly different angle of incidence. The symbols of HS, $F_{XUV}$, $F_{IR}$, $PD_0$, $PD_{IR}$, $PD_{XUV}$, $i_{IR}$, $i_{XUV}$, $i_0$, $n_{IR}$ and $n_{XUV}$, are the same as those used in Figure 1. The $PD_{XUV}$ is placed on a translation stage (TS) in order to be able to move it out of the XUV beam path and record the XUV spectrum by a conventional XUV spectrometer (Spec.). Ap.: An aperture serving as spatial filter in order to optimize the transmission/detection of the RHH. The polarizer after the HS plate is used in order to block the IR photons of the reflected by the solid s-pol. pre-pulse. In this arrangement together with $i_{IR}$, $i_{XUV}$ and $i_0$ photocurrents, the variations of the CEP ($S_{CEP}$) and target position ($S_t$) are recorded for each laser shot. (**b**) A simplified calculation (performed for visualization reasons) of the multi-peak structure of $P_n$ which corresponds to the number of IR photon absorbed towards RHHG. The distribution has been calculated by considering a sum of Gaussian peaks, using as relative amplitudes and widths, the amplitudes and the widths of the harmonics shown in Figure 3b. Also we have considered that, $N_0 \sim 10^{16}$, $\eta_{XUV} \sim 10^{-3}$–$10^{-2}$, an IR attenuation factor of $10^8$, a negligible contribution of the IR uncorrelated points, and that the total number of points along the anticorrelation diagonal is $10^3$. The numbers in blue show the numbers of points contributing in each peak of the $P_n$. The peak corresponding to the 17th and 16th harmonics (and maybe 15th), practically cannot be resolved because they contain 1 and 0.5 points, respectively.

The photon number of the main IR pulse after the interaction and the photon number of the generated XUV radiation are measured by the $PD_{IR}$ (providing the photocurrent value $i_{IR}$) and the $PD_{XUV}$ (providing the photocurrent value $i_{XUV}$), respectively. An aperture (Ap.) placed before the $PD_{XUV}$ will serve for spatial filtering in order to optimize the transmission/detection of the RHH. The $PD_{XUV}$ will be placed on a translation stage providing the ability to insert it in or out from the XUV beam path and record the harmonic spectrum by means of a conventional spectrometer. In order to block the IR photons of the pre-pulse after the interaction, a polarizer allowing the p-polarized IR photons to pass through will be placed before the $PD_{IR}$. Additionally, CEP variations ($S_{CEP}$) and the values of the target position ($S_t$) will be recorded for each laser shot in order to be able to discard from the data set any potential shots depicting strong instabilities.

Considering that the photon number of the main pulse is $N_0 \approx 10^{16}$ photons per pulse for $a_0 \approx 2$ the non-linearity of the ROM harmonics generation process is ~6 (see Figure 3c), the energy stability of the main pulse ≈1%, the number of available shots per target ≈$10^6$, it is estimated that the joint IR-vs-XUV distribution (($i_{IR}$, $i_{XUV}$) plot in Figure 2a) contains $k \approx 10^6$ points. This results to an anti-correlation diagonal of width $w_{anti} = W_{joint}/k^{\frac{1}{2}} \sim 6 \times 10^{-3}\%$. According to the relation $N_q^{(XUV)} \gtrsim N_{IR} \cdot w_{anti}/A$ provided in Section 1, and considering that $N_{IR} \approx N_0$ and the XUV absorption in the plasma is negligible i.e., $A = 1$, it is estimated the QS can resolve a background free multi-peak structure of $P_n$ when the outgoing from the medium XUV photon number is $N_q^{(XUV)} \gtrsim 6 \times 10^{11}$ photons per pulse. This harmonic photon number is well below the estimated photon number of the RHHG harmonics ($10^{12}$–$10^{13}$ photons per RHHG harmonic), justifying the feasibility of the experiment. However, the use of a $w_{\text{anti}} \sim 6 \times 10^{-3}$% results in an anticorrelation diagonal which contains $\sim 2 \times 10^2$ points, which may reduce the visibility of the multi-peak structure of $P_n$. As has been discussed in Section 1, this can be mitigated by increasing the width of the anticorrelation diagonal by a factor of $\sim 5$, i.e., $w_{\text{anti}}^{(opt.)} \sim 3 \times 10^{-2}$%. Considering that $N_q^{(XUV)} \approx 5 \times 10^{12}$, this value of $w_{\text{anti}}^{(opt.)}$ satisfies the condition $w_{\text{anti}} < w_{\text{anti}}^{(opt.)} < N_q^{(XUV)}/N_{IR}$ and increases the number of points to $\sim 10^3$ thus improving the statistical significance of $P_n$ (Figure 4b). We note that the dynamic range of the detector should be $D \gtrsim \frac{N_{IR}}{N_q^{(XUV)}} = \frac{10^{16}}{5 \times 10^{12}} \sim 2 \times 10^3$ i.e., $D > 12$ bit. In the case that $N_q^{(XUV)} \approx 10^{12}$ (corresponding to $\eta_{\text{XUV}} \sim 10^{-4}$), the multi-peak structure of $P_n$ (build on the top of a small background IR photon number distribution as $w_{\text{anti}}^{(opt.)} > N_q^{(XUV)}/N_{IR}$) can be recorded by means of a 14-bit detection system.

Here, it is worth stressing that although the study has been conducted for RHH (having a spectrum with well confined harmonic peaks) generated by the interaction of the solid surface with a sinusoidal (CEP = $\pi/2$) few-cycle driving field, it is applicable for any CEP value including those leading to the emission of an XUV radiation of continuous spectrum. The latter will result in a $P_n$ which depicts a continuum photon number distribution. Investigations concerning the dependence of $P_n$ on the CEP of the driving field, can be considered of particular importance as they can provide access to the sub-cycle relativistic quantum electrodynamics of the interaction.

Finally, we note that the above estimations take into account energy conservation and the operation principle of QS as was implemented in the HHG process in atoms [26,27] and semiconductors [28], having utilized fully quantized approaches. In these studies, $P_n$ was obtained considering the quantum nature of the electromagnetic radiation as well as the back-action of the HHG process into the quantum state of the driving IR laser field. Consequently, in the case of relativistic laser-plasma interaction and RHHG, the theoretical description of $P_n$ requires the development of a fully quantized theoretical model which incorporates the quantized nature of the driving IR field (coherent light states) into RHHG models [32,39,40], which is traditionally used for the description of laser–plasma interaction. A suitable platform to incorporate the quantum nature of the IR field into the theoretical description is likely to be provided by a separation into a classical contribution (representing the expectation value of the field) and a quantum correction, as was done for HHG in gas-phase targets [26], with the RHH dynamics driven by the classical contribution; the RHH dynamics can then be fed back into the quantum state of the IR field through their coupling with the quantum-correction operator. In contrast to the gas-phase HHG, however, the particle-in-cell (PIC) nature of simulations in order to accurately describe the RHHG models pose additional challenges to incorporate the response back to the quantum state of the driving field. As such, these theory developments present formidable challenges, but they are extremely important as they will provide access to investigations of relativistic quantum electrodynamics and the generation of non-classical light states, similar to the lines of recent investigations in gas phase HHG [26].

## 4. Conclusions

In this work we propose the extension of the quantum spectrometer method to relativistic interactions using the laser–plasma RHHG process. In particular, we highlight that the unique properties of the RHH source reported in reference [29] (high repetition rate, stability of the laser-plasma interaction, number of shots per target), and the predicted high conversion efficiency for the generation of the RHH, fulfill all the requirements of the quantum spectrometer method ensuring its proper operation. Specifically, we show that by using a stable laser-plasma high harmonic source, which delivers RHH with conver-sion efficiency of in the range of $10^{-3}$–$10^{-2}$ (in the ~17 eV spectral region), the harmonic spectrum can be obtained by the photon number distribution of the IR field exiting the HHG process when the quantum spectrometer accumulates ~$10^6$ shots. The successful realization of this proposal will be a first vital step towards quantum optical studies in relativistic laser–plasma interactions. High power CEP-stabilized few-cycle laser driven surface plasma-based harmonic sources coming up in a large-scale facility like ELI-ALPS [52] will further serve such investigations and in conjunction with HHG determine also the role of CEP effects.

**Author Contributions:** T.L., R.L.-M., S.H.: contributed to the data analysis and the design of the proposed experiment; S.K., I.L., J.R.-D., P.S., E.P., M.F.C., M.L.: contributed to the data analysis and validation; P.T.: Conceived, supervised and contributed to all aspects of the work. All authors contributed to the manuscript preparation. All authors have read and agreed to the published version of the manuscript.

**Funding:** P.T. group acknowledges LASERLABEUROPE (EU's Horizon 2020 Grant No. 871124), FORTH Synergy Grant AgiIDA, HELLAS-CH (MIS Grant No. 5002735) [which is implemented under the Action for Strengthening Research and Innovation Infrastructures, funded by the Operational Program Competitiveness, Entrepreneurship and Innovation (NSRF 20142020) and co-financed by Greece and the European Union (European Regional Development Fund)], and the European Unions Horizon 2020 research. ELI-ALPS is supported by the European Union and co-financed by the European Regional Development Fund (GINOP Grant No. 2.3.6-15-2015-00001). R.L.-M and S.H. acknowledges financial support from the European Research Council (ERC Advanced Grant ExCoMet 694596) and LASERLAB-EUROPE (H2020-EU.1.4.1.2. grant agreement ID 654148). S.K. acknowledges project no. 2018-2.1.14-TÉT-CN-2018-00040 which has been implemented with the support provided from the National Research, Development and Innovation Fund of Hungary, financed under the 2018-2.1.14-TÉT-CN funding scheme. The ELI-ALPS project (GINOP-2.3.6-15-2015-00001) is supported by the European Union and co-financed by the European Regional Development Fund. M.L. group acknowledges the European Research Council (ERC AdG) NOQIA, Spanish Ministry MINECO and State Research Agency AEI (FIDEUA PID2019-106901GB-I00/10.13039/501100011033, SEVERO OCHOA No. SEV-2015-0522 and CEX2019-000910-S, FPI), European Social Fund, Fundaci Cellex, Fundaci Mir-Puig, Generalitat de Catalunya (AGAUR Grant No. 2017 SGR 1341, CERCA program, QuantumCAT U16-011424, co-funded by ERDF Operational Program of Catalonia 2014-2020), MINECO-EU QUANTERA MAQS (funded by State Research Agency (AEI) PCI2019-111828-2/10.13039/501100011033), EU Horizon 2020 FETOPEN OPTOLogic (Grant No 899794), and the National Science Centre, Poland-Symfonia Grant No. 2016/20/W/ST4/00314, Marie Sklodowska-Curie grant STRETCH No 101029393. M.F.C. acknowledges the Czech Science Foundation (GACR) (Grant number: 20-24805J). J.R.-D. acknowledges the Secretaria d'Universitats i Recerca del Departament d'Empresa i Coneixement de la Generalitat de Catalunya, as well as the European Social Fund (L'FSE inverteix en el teu futur)–FEDER.

**Institutional Review Board Statement:** Not applicable.

**Informed Consent Statement:** Not applicable.

**Data Availability Statement:** The data presented in this study are available upon request from the corresponding author.

**Acknowledgments:** We acknowledge the projects listed in the Funding section.

**Conflicts of Interest:** The authors declare no conflict of interest.